# Modeling Power Angle Spectrum and Antenna Pattern Directions in Multipath Propagation Environment

Jan M. Kelner and Cezary Ziółkowski
Institute of Telecommunications, Faculty of Electronics, Military University of Technology, Warsaw, Poland
{jan.kelner, cezary.ziolkowski}@wat.edu.pl

*Abstract*—Most propagation models do not consider the influence of antenna patterns on the parameters and characteristics of received signals. This assumption is equivalent to the use of isotropic or omnidirectional antennas in these models. Empirical measurement results indicate that the radiation pattern, gain and direction of directional antennas significantly influence on properties of the received signal. This fact shows that consideration the directional antennas in propagation models is very important especially in the context of emerging telecommunication technologies such as beamforming or massive MIMO. The purpose of this paper is to present the modeling method of power angular spectrum and direction of antenna patterns in a multipath propagation environment.

## I. Introduction

A power angle spectrum (PAS) is the basic characteristic used to estimate and model an angular dispersion in wireless channels. PAS can be determined in azimuth and elevation planes. If the model includes both planes, then we are talking about 3D models [1],[2],[3]. In 2D [4],[5],[6] models, only the azimuth plane is considered, and then the PAS acronym means the power azimuth spectrum.

PAS is a practical channel characteristic used primarily in empirical measurements. In the case of radio channel modeling, theoretical characteristic, i.e., a probability density function (PDF) of angle of arrival (AOA), is more commonly used.

If the channel model does not consider a pattern of a receiver (Rx) antenna, then we can assume that an omnidirectional (for a 2D model) or isotropic antenna is in this model, e.g., [4],[5]. In this case, PAS and PDF of AOA shapes are identical, while the total power of a received signal is proportionality factor between these characteristics [7].

If the receiving antenna is directional, then we should distinguish two types of theoretical characteristics, i.e., PDF of AOA seen at the Rx antenna input and PDF of angle of reception (AOR) seen from the Rx input, i.e., the Rx antenna output [3]. Therefore, PDF of AOR and PAS include the receiving antenna pattern.

Two approaches are used for PAS and PDF of AOA/AOR modeling. The first is the geometric channel model (GCM), which defines the area of scatterer location of signal components and their density, e.g., [1],[4]. The second is empirical models, which are based on standard statistical distributions such as Laplacian or Gaussian [6]. This model type is also used to describe PAS in standard models, so-called reference models, such as the 3GPP model [8]. There are also mixed models, e.g. [2],[3],[5], where delayed scatter components of the received signal are defined geometrically, while local scattering components are statistical in nature.

The aim of the paper is to show the impact of the directional antenna patterns and their directions on the PAS in the azimuth plane. In this case, the multi-elliptical channel model (MCM) is used for modeling [2],[3]. The presented results show a significant influence of the antenna patterns and their directions on the spatial dispersion and total power of the received signal. The analysis of this issue is particularly important for future telecommunication systems, especially in the context of beamforming, massive MIMO technology, green communications, and 5G systems.

This paper is organized as follows. In Section II, modeling method of PAS and antenna pattern directions using MCM is briefly described. Evaluation of the influence of directional antennas on PAS is presented in Section III. Section IV contains conclusions.

## II. Modeling PAS and Antenna Pattern

In MCM, three types of the received signal components are present. These are the direct path component, local scattering components occurring near the receiving antenna, and the delayed scattering components. To describe the local scattering, the von Mises distribution is used. The scatterers of delayed components are located on coaxial ellipses (2D) [5],[7],[9] or half-ellipsoids (3D) [2],[3]. In this paper, the PAS analysis concerns only the azimuth plane, so the following description refers only to ellipses, even in regard to literature describing the 3D model. The transmitter (Tx) and Rx are located in the foci of the ellipses. The MCM geometry is shown in Fig. 1.

The basic input data for MCM is the power delay profile (PDP), which represents the power distribution on time clusters. PDP delays define sizes of the ellipses, while PDP powers describe the powers of individual signal components [2],[3],[5],[7],[9].

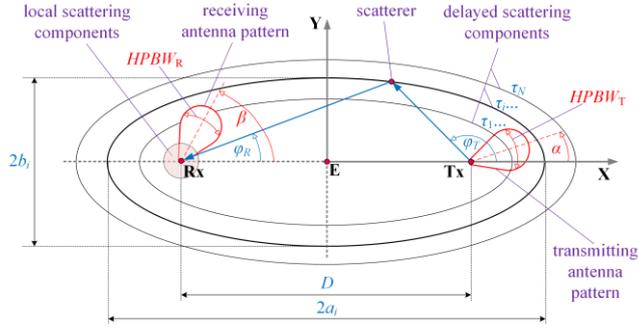

Fig. 1. Geometry of MCM.

The transmitting antenna pattern is used to generate angles of departure (AOD) [2],[9]. Thus, the shape, half-power beam width (HPBW), $HPBW_T$, and $\alpha$ direction of the Tx antenna pattern has a significant impact on the PDF of AOD, and indirectly also on the PDF of AOA. In the general case, any shape of the antenna pattern may be used. In this paper, we use the Gaussian to model the main lobe of the directional antennas [10]

$$g_T(\varphi_T) = \exp\left(-4\ln 2 \frac{(\varphi_T - \alpha)^2}{HPBW_T^2}\right) \quad (1)$$

Thus, AODs, $\varphi_T$, are generated on the basis of the PDF defined as [2],[9]

$$f_T(\varphi_T) = C_T(HPBW_T) g_T^2(\varphi_T) \quad (2)$$

where $C_T$ is constant normalizing the PDF of AOD that depends on $HPBW_T$.

For the delayed scattering components, AOAs, $\varphi_R$, are determined for the $i$th ellipse and each generated AOD [9]

$$\varphi_{R,i} = \text{sgn}(\varphi_T)\arccos\frac{2e_i + (1+e_i^2)\cos(\varphi_T)}{1+e_i^2 + 2e_i\cos(\varphi_T)} \quad (3)$$

where $e_i = d/(d+c\tau_i)$ is the eccentricity of the $i$th ellipse defined by $\tau_i$ delay for $i=1,2,...,N$, $N+1$ is the number of time clusters occurred in the PDP, $d$ is the Tx-Rx distance, and c is the speed of light.

For the delayed scattering components, PDFs of AOA, $f_{d,i}(\varphi_R)$, are determined from the obtained sets of AOAs for each ellipse. These PDFs can be determined in two ways, i.e., based on histograms of AOA [5] or component powers [2],[9]. Conducted studies show that both approaches are equivalent.

For the local scattering components, AOAs are generated based on the von Mises distribution

$$f_{l,0}(\varphi_R) = \frac{\exp(\mu\cos\varphi_R)}{2\pi I_0(\mu)} \quad (4)$$

where $I_0(\cdot)$ is the zero-order modified Bessel function and $\mu$ is the distribution parameter describing the intensity of the local scattering.

The resulting PDF of AOA seen at the receiving antenna input is described by [2],[9]

$$f(\varphi_R) = \sum_{i=1}^{N}\frac{P_i}{P_R}f_{d,i}(\varphi_R) + \frac{1}{\kappa+1}\frac{P_0}{P_R}f_{l,0}(\varphi_R) + \frac{\kappa}{\kappa+1}\frac{P_0}{P_R}\delta(\varphi_R) \quad (5)$$

where $\kappa$ is the Rician factor, $P_i$ is the power of the $i$th PDP time cluster corresponding to the delay $\tau_i$ for $i=0,1,...,N$, $P_R = \sum_{i=0}^{N}P_i$ is the total power of the received signal, and $\delta(\cdot)$ is the Dirac delta function.

In the case of the isotropic or omnidirectional antennas, the component powers of the received signal are generated on the basis of uniform distributions defined for each clusters. For delayed and local scattering components, we have respectively [2],[5],[9]

$$f_i(p_{ij}) = \begin{cases} M_i/(2P_i) & \text{for } p_{ij} \in \langle 0, 2P_i/M_i\rangle \\ 0 & \text{for } p_{ij} \notin \langle 0, 2P_i/M_i\rangle \end{cases} \quad (6)$$

$$f_i(p_{0j}) = \begin{cases} (1+\kappa)M_0/(2P_0) \\ \quad \text{for } p_{0j} \in \langle 0, 2P_i/((1+\kappa)M_0)\rangle \\ 0 \\ \quad \text{for } p_{0j} \notin \langle 0, 2P_i/((1+\kappa)M_0)\rangle \end{cases} \quad (7)$$

where $p_{ij}$ is the power of the $j$th path of the $i$th cluster seen at the Rx antenna input, $M_i$ is the number of generated paths in the $i$th cluster, $i=0,1,...,N$, and $j=1,2,...,M_i$.

In the case of the directional receiving antenna, PAS and PDF of AOD consider also its pattern. The pattern of this antenna can be described analogously to the transmitting antenna, i.e.,

$$g_R(\varphi_R) = \exp\left(-4\ln 2 \frac{(\varphi_R - \beta)^2}{HPBW_R^2}\right) \quad (8)$$

where $\beta$ and $HPBW_R$ are the direction and width of the main lobe of the receiving antenna pattern, respectively.

The receiving antenna pattern is used to determine the power of each path seen at the Rx antenna output

$$p_{R,ij} = p_{ij} g_R^2(\varphi_{R,ij}) \quad (9)$$

where $p_{R,ij}$ is the power of the $ij$th path seen at the Rx input, corresponds to AOA/AOR, $\varphi_{R,ij}$.

The obtained powers are used to determine the estimator of the PDF of AOD that is seen at the Rx input. For this purpose, we use the histogram method for the path powers [3]

$$\tilde{f}_R(\varphi_R) = C_R \frac{\sum_{\mathbf{L}(\varphi_R)} p_{R,ij}(\varphi_{R,ij})}{\sum_{i=0}^{N}\sum_{j=1}^{M_i} p_{R,ij}(\varphi_{R,ij})} \quad (10)$$

where $\mathbf{L}(\varphi_R) = \{(i,j): \varphi_{R,ij} \in (\varphi_R \pm \varepsilon_\varphi)\}$, $C_R$ is constant normalizing the PDF of AOR, and $\varepsilon_\varphi$ is a neighborhood of $\varphi_R$.

Then, we can describe the PAS seen at the input Rx as [3],[7]

$$P(\varphi_R) = P_R \tilde{f}_R(\varphi_R) \quad (11)$$

The PAS modeling method described above is used in the studies presented in this paper.

## III. IMPACT OF ANTENNA PATTERNS ON PAS

The impact evaluation of the antenna patterns on PAS is carried out for the urban macro (UMa) scenario and non-line-of-sight (NLOS) conditions, $\kappa = 0$. For this aim, we use PDP based on the 3GPP tapped delay line (TDL) model, i.e., TDL-B [8, Table 7.7.2-2] for the frequency range, $f_c = 2.4 - 2.7$ GHz, and rms delay spread, $DS = 363$ ns [8, Table 7.7.3-2]. Three types of antennas are analyzed, i.e., corner reflector (CR), parabolic grid (PG), and base station antenna (BS). The HPBWs of these antennas are $HPBW_{CR} = 58°$, $HPBW_{PG} = 10°$, and $HPBW_{BS} = 68°$ for CR, PG [11], and BS (Kathrein 80010715) [12], respectively. The Tx-Rx distances are $d = 800$ m and $d = 400$ m for BS and CR or PG, respectively.

Figures 2, 4, and 6 show PASs for $\alpha = 180°$, selected $\beta$, and three analyzed types of antennas, i.e., BS, CR, and PG, respectively. While, Figs. 3, 5, and 7 present PASs for $\beta = 0°$, selected $\alpha$, and three antenna types, i.e., BS, CR, and PG, respectively.

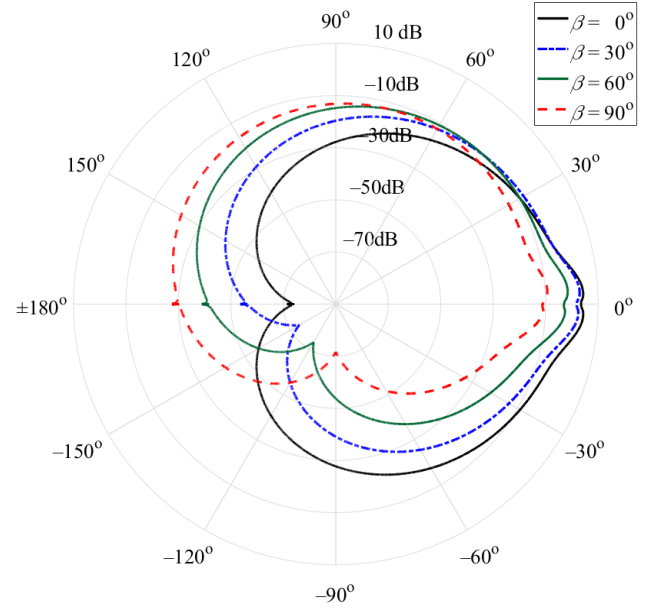

Fig. 2. PASs for BS, $\alpha = 180°$ and selected $\beta$.

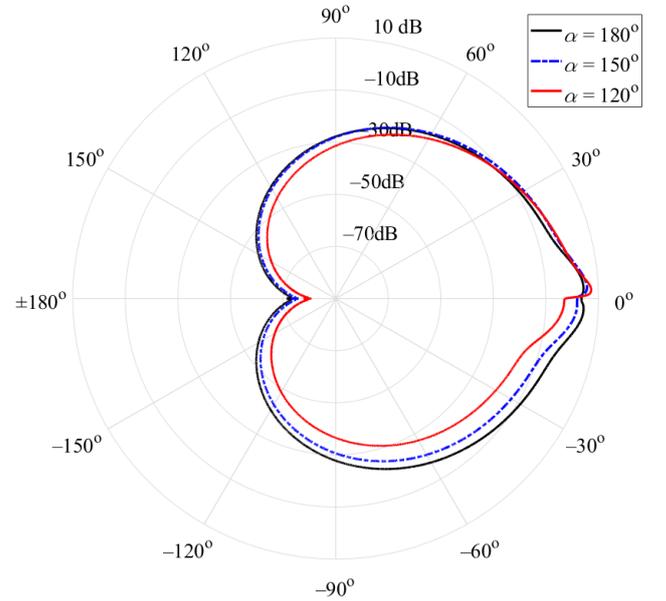

Fig. 3. PASs for BS, $\beta = 0°$ and selected $\alpha$.

The obtained results show that PASs are symmetrical for cases the antennas are directed at themselves, i.e., for $\alpha = 180°$ and $\beta = 0°$. Changing the direction of one antenna results a deformation of the PAS symmetry.

For BS and CR, PAS shapes are similar despite different distances. For BS and CR, HPBWs are approximate. Thus,

this parameter of the antennas have a significant influence on the PAS shape. For very selective PG, PASs are also selective.

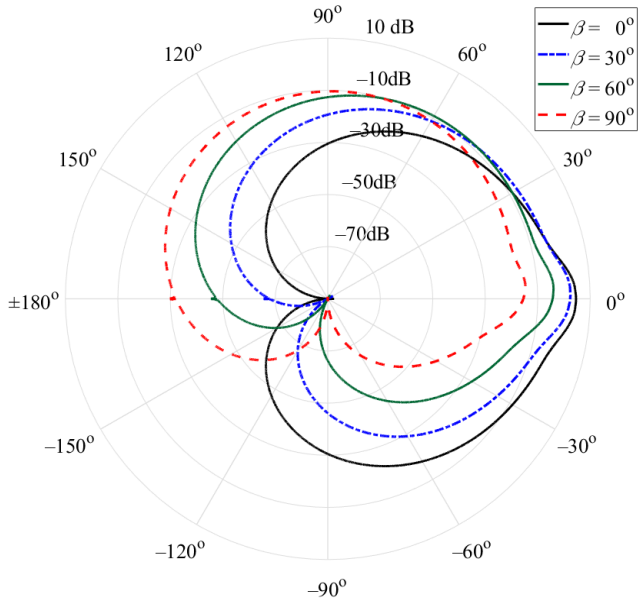

Fig. 4. PASs for CR, $\alpha = 180°$ and selected $\beta$.

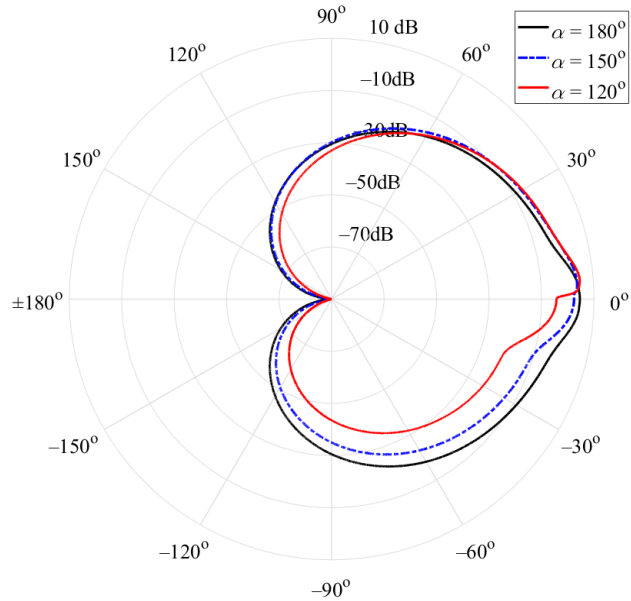

Fig. 5. PASs for CR, $\beta = 0°$ and selected $\alpha$.

For $\alpha = 180°$, changes of $\beta$ cause major changes in the shape and orientation of the received PAS. For $\beta = 0°$, changes of $\alpha$ cause relatively minor changes of the PAS. Thus, impact of changing the receiving antenna direction on the PAS is more significantly.

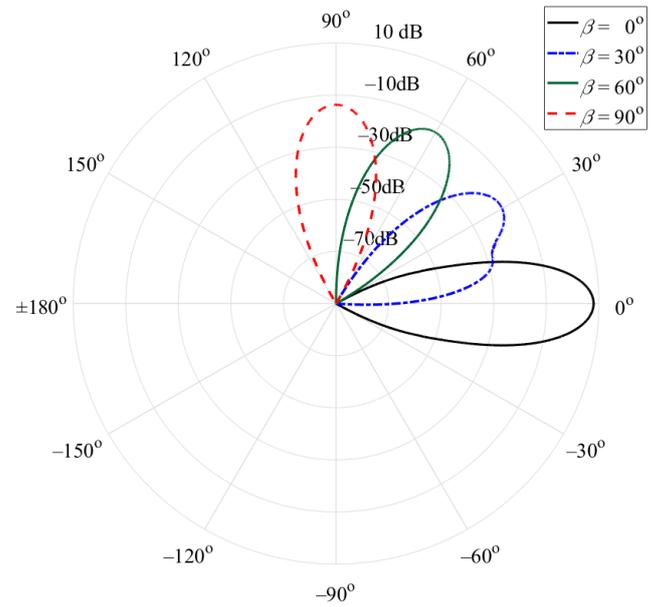

Fig. 6. PASs for PG, $\alpha = 180°$ and selected $\beta$.

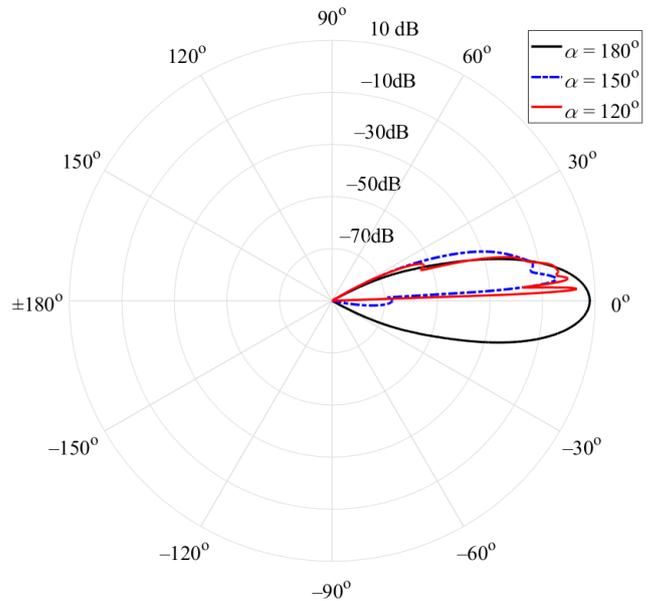

Fig. 7. PASs for PG, $\beta = 0°$ and selected $\alpha$.

The obtained graphs show that the proper orientations of the transmitting and receiving antennas significantly impact on the PAS shape, angular dispersion, and power of the received signal.

Presented approach for the PAS modeling is used in [13] to modify a path loss model for various orientation of directional antennas. This path loss model only for antennas directed at each other is shown in [11].

IV. CONCLUSION

In this paper, the authors presented the methodology of the PAS modeling, which considers the directional antennas.

The obtained simulation results show that the PAS shape closely depends on the HPBWs and directions of the antenna patterns. The proposed methodology can be used to assess the impact of interference between antenna systems for their different orientations. This is important in an electromagnetic compatibility evaluation of radiocommunication systems. The modeling method of the directional antenna and PAS can also be used to modify propagation models that do not consider the antenna patterns.